
\input harvmac
 %
\catcode`@=11   
 %

 %

\def\intem#1{\par\leavevmode%
              \llap{\hbox to\parindent{{#1}\hfill}}\ignorespaces}
 %
\def\niche#1#2{\setbox0=\hbox{#2}\hangindent\wd0\hangafter-#1%
                \par\noindent\hskip-\wd0\hbox to\wd0{\box0}}

 %

 %
\let\ii=\i          
\def\,{\hskip1.5pt}
 %
 %
\let\a=\alpha
\let\b=\beta
\let\c=\chi
\let\d=\delta       \let\vd=\partial             
     \let\ve=\varepsilon
\let\f=\phi         \let\vf=\varphi              
\let\g=\gamma                                    

\let\i=\iota
\let\j=\psi                                      

\let\l=\lambda                                   \let\L=\Lambda
\let\m=\mu

\let\q=\theta       \let\vq=\vartheta            \let\Q=\Theta
         
\let\s=\sigma       \let\vs=\varsigma            \let\S=\Sigma

\let\w=\omega

 %

\def\con{\ifmmode \hbox{\bf*} \else{\bf*}\fi}   
\def\bo{{\raise.15ex\hbox{\large$\Box\kern-.39em$}}}

\let\into=\hookrightarrow

\def\dual{\relax\leavevmode\lower.9ex\hbox{\titlerms*}}
\def\define{\buildrel\rm def\over =}
\let\id=\equiv
\let\8=\otimes
 %

 %

\let\2=\underline

\let\Tw=\widetilde

\font\eightrm=cmr8
\def\6(#1){\relax\leavevmode\hbox{\eightrm(}#1\hbox{\eightrm)}}
\def\BM#1{\relax\leavevmode\setbox0=\hbox{$#1$}
           \kern-.025em\copy0\kern-\wd0
            \kern.05em\copy0\kern-\wd0
             \kern-.025em\raise.0433em\box0}

\def\0#1{\relax\ifmmode\mathaccent"7017{#1}     
                \else\accent23#1\relax\fi}      
\def\7#1#2{{\mathop{\null#2}\limits^{#1}}}      
\def\5#1#2{{\mathop{\null#2}\limits_{#1}}}      


     %
\newbox\t@b@x
\def\rightarrowfill{$\m@th \mathord- \mkern-6mu
     \cleaders\hbox{$\mkern-2mu \mathord- \mkern-2mu$}\hfill
      \mkern-6mu \mathord\rightarrow$}
\def\tooo#1{\setbox\t@b@x=\hbox{$\scriptstyle#1$}%
             \mathrel{\mathop{\hbox to\wd\t@b@x{\rightarrowfill}}%
              \limits^{#1}}\,}
\def\leftarrowfill{$\m@th \mathord\leftarrow \mkern-6mu
     \cleaders\hbox{$\mkern-2mu \mathord- \mkern-2mu$}\hfill
      \mkern-6mu \mathord-$}
\def\froo#1{\setbox\t@b@x=\hbox{$\scriptstyle#1$}%
             \mathrel{\mathop{\hbox to\wd\t@b@x{\leftarrowfill}}%
              \limits^{#1}}\,}
\def\frac#1#2{{#1\over#2}}
\def\frc#1#2{\relax\ifmmode{\textstyle\frac{#1}{#2}}
                    \else$\frac{#1}{#2}$\fi}        
\def\inv#1{\frc{1}{#1}}                             
 %
\newskip\humongous \humongous=0pt plus 1000pt minus 1000pt
\def\caja{\mathsurround=0pt}
\newif\ifdtup
\def\panorama{\global\dtuptrue \openup2\jot \caja
     \everycr{\noalign{\ifdtup \global\dtupfalse
      \vskip-\lineskiplimit \vskip\normallineskiplimit
      \else \penalty\interdisplaylinepenalty \fi}}}
\def\eqalign#1{\,\vcenter{\openup2\jot \caja
     \ialign{\strut \hfil$\displaystyle{##}$&$
      \displaystyle{{}##}$\hfil\crcr#1\crcr}}\,}
\def\eqalignno#1{\panorama \tabskip=\humongous
     \halign to\displaywidth{\hfil$\displaystyle{##}$
      \tabskip=0pt&$\displaystyle{{}##}$\hfil
       \tabskip=\humongous&\llap{$##$}\tabskip=0pt\crcr#1\crcr}}

 %

 %
\def\inbar{\vrule height1.5ex width.04em depth-.04ex}
\def\sinbar{\vrule height1ex width.03em depth-.02ex}
\def\ssinbar{\vrule height.7ex width.025em depth-.01ex}
\font\cmss=cmss10
\font\cmsss=cmss10 at 7pt
\def\ZZ{\relax\leavevmode
               \ifmmode\mathchoice
                      {\hbox{\cmss Z\kern-.4em Z}}
                      {\hbox{\cmss Z\kern-.4em Z}}
                      {\lower.9pt\hbox{\cmsss Z\kern-.36em Z}}
                      {\lower1.2pt\hbox{\cmsss Z\kern-.36em Z}}
               \else{\cmss Z\kern-.4em Z}\fi}
\def\Ik{\relax{\rm I\kern-.18em k}}
\def\IC{\relax\leavevmode
               \ifmmode\mathchoice
                      {\hbox{\kern.33em\inbar\kern-.35em{\rm C}}}
                      {\hbox{\kern.33em\inbar\kern-.35em{\rm C}}}
                      {\hbox{\kern.27em\sinbar\kern-.29em{\sevenrm C}}}
                      {\hbox{\kern.24em\ssinbar\kern-.26em{\fiverm C}}}
               \else{\hbox{\kern.3em\inbar\kern-.32em{\rm C}}}\fi}
\def\IP{\relax{\rm I\kern-.18em P}}
\def\IR{\relax{\rm I\kern-.18em R}}
\def\Ione{\relax{\rm 1\kern-3pt l}}

 %

\def\CP#1{\relax\ifmmode\IP^{#1}\else\IP$^{#1}$\fi}
\def\cP#1{\relax\ifmmode\IC{\rm P}^{#1}\else\IC${\rm P}^{#1}$\fi}
\def\WP#1#2{\relax\ifmmode\IP^{#1}_{#2}\else\IP$^{#1}_{#2}$\fi}
\def\K#1#2{\relax\def\normalbaselines{\baselineskip12pt\lineskip3pt
                                       \lineskiplimit3pt}
        \left[\matrix{#1}\right.\!\left\|\,\matrix{#2}\right]}
\def\muthstrut{\vphantom1}
\def\mutrix#1{\null\,\vcenter{\normalbaselines\m@th
        \ialign{\hfil$##$\hfil&&~\hfil$##$\hfill\crcr
            \muthstrut\crcr\noalign{\kern-\baselineskip}
            #1\crcr\muthstrut\crcr\noalign{\kern-\baselineskip}}}\,}

\def\EU{\relax\ifmmode \c_{{}_E} \else$\c_{{}_E}$\fi}
\def\TM{\relax\ifmmode {\cal T_M} \else$\cal T_M$\fi}
\def\TW{\relax\ifmmode {\cal T_W} \else$\cal T_W$\fi}
\def\CM{\relax\ifmmode {\cal T\rlap{\bf*}\!\!_M}
               \else$\cal T\rlap{\bf*}\!\!_M$\fi}
\def\hm#1#2{\relax\ifmmode H^{#1}({\cal M},{#2})
                   \else$H^{#1}({\cal M},{#2})$\fi}
\def\CP#1{\relax\ifmmode\IP^{#1}\else\IP$^{#1}$\fi}
\def\cP#1{\relax\ifmmode\IC{\rm P}^{#1}\else$\IC{\rm P}^{#1}$\fi}
\def\sll#1{\rlap{\,\raise1pt\hbox{/}}{#1}}

\let\ddd=\displaystyle
%

 %

\def\CY{Calabi-\kern-.2em Yau}
\def\?{d\kern-.3em\raise.64ex\hbox{-}}   
\def\9{\raise.43ex\hbox{-}\kern-.37em D} 
\def\3{\ifmmode\ldots\else$\ldots$\fi}
\def\ping{\nobreak\par\centerline{---$\circ$---}\goodbreak\bigskip}

 %

\def\I#1{{\it ibid.\,}{\bf#1\,}}

\def\NP#1{{\it Nucl.\,Phys.\,}{\bf#1\,}}
\def\PL#1{{\it Phys.\,Lett.\,}{\bf#1\,}}

\def\Pre#1{{\it #1\ University report}}

\def\MPL#1{{\it Mod.\,Phys.\,Lett.\,}{\bf#1\,}}

\def\CMP#1{{\it Commun.\,Math.\,Phys.\,}{\bf#1\,}}
\def\CQG#1{{\it Class.\,Quant.\,Grav.\,}{\bf#1\,}}
\def\IJMP#1{{\it Int.\,J.\,Mod.\,Phys.\,}{\bf#1\,}}

 %
\catcode`@=12                   

     %
 
 \def\pmod#1{\allowbreak\mkern8mu({\rm mod}\,\,#1)}
 \def\Jb{\skew6\bar{J}}
 \def\LGO{Landau-Ginzburg orbifold}
 \def\Xb{\relax\leavevmode\hbox{$X$\kern-.6em%
                   \vrule height.4pt width5.7pt depth-1.8ex}\kern1pt}
 \def\rd{{\rm d}}
 \def\ping{\par\centerline{---$\circ$---}\par}

 \def\eqalignthree#1{\,\vcenter{\openup2\jot \caja
     \ialign{\strut \hfil$\displaystyle{##}$&
                         $\displaystyle{{}##}$\hfil&
                         $\displaystyle{{}##}$\hfil&
                         $\displaystyle{{}##}$\hfil\crcr#1\crcr}}\,}

     %

\baselineskip=13.6 pt plus 2pt minus1pt


\noblackbox

\nopagenumbers\abstractfont\hsize=\hstitle
\null\vskip-40pt
\rightline{\vbox{\baselineskip12pt\hbox{CERN-TH-6381/92}
                                  \hbox{HUTMP-91/B315}
                                  \hbox{UTTG-21-91}}}%
\vfill
\centerline{\titlefont Classical {\it vs.}\ Landau-Ginzburg}
\vskip10pt
\centerline{\titlefont Geometry of Compactification}
\abstractfont\vfill\pageno=0

\centerline{Per Berglund}                                \vskip-.2ex
 \centerline{\it Theory Division, CERN}                  \vskip-.4ex
 \centerline{\it CH-1211 Geneva 23, Switzerland}         \vskip-.4ex
 \centerline{and}                                        \vskip-.4ex
 \centerline{\it Theory Group, Physics Department}       \vskip-.4ex
 \centerline{\it University of Texas, Austin, TX 78712}
\vskip .1in
\centerline{Brian R.~Greene}                                \vskip-.2ex
 \centerline{\it Physics Department, Cornell University} \vskip-.4ex
 \centerline{\it Ithaca, NY 14853}                       \vskip0ex
\vskip .05in
\centerline{and}
\vskip .05in
\centerline{Tristan H\"ubsch\footnote{$^{\spadesuit}$}
      {On leave from Institute ``Rudjer Bo\v skovi\'c'',
       Zagreb, Croatia.}}                                \vskip-.2ex
 \centerline{\it Departments of Mathematics and Physics} \vskip-.4ex
 \centerline{\it Harvard University, Cambridge, MA~02138}\vskip-.4ex
\vfill

\centerline{ABSTRACT}\nobreak\vglue.8ex
\vbox{\narrower\baselineskip=12pt
We consider superstring compactifications where both the classical
description, in terms of a \CY\ manifold, and also the quantum theory
is known in terms of a \LGO\ model.  In particular, we study (smooth)
Calabi-Yau examples in which there are obstructions to parametrizing
all of the complex structure cohomology by polynomial deformations
thus requiring the analysis based on exact and spectral sequences.
General arguments ensure that the Landau-Ginzburg chiral ring copes
with such a situation by having a nontrivial contribution from
twisted sectors. Beyond the expected final agreement between the
mathematical and physical approaches, we find a direct correspondence
between the analysis of each, thus giving a more complete mathematical
understanding of twisted sectors. Furthermore, this approach shows
that physical reasoning based upon spectral flow arguments for
determining the spectrum of \LGO\ models finds direct mathematical
justification in Koszul complex calculations and also that careful
point-field analysis continues to recover surprisingly much of the
stringy features.}

\Date{\vbox{\line{CERN-TH-6381/92\hfill}
            \line{2/\number\yearltd \hfill}}}

\nref\rPDM{P.~Green and T.~H\"ubsch: \CMP{113}(1987)505.}

\nref\rDoron{D.~Gepner: \PL{199B}(1987)380.}

\nref\rGVW{B.R.~Greene, C.~Vafa and N.P.~Warner: \NP{B324}(1989)371.}

\nref\rMartinec{E.~Martinec: \PL{171B}(1986)189.}

\nref\rLGO{C.~Vafa: \MPL{A4}(1989)1169\semi
      K.~Intrilligator and C.~Vafa: \NP{B339}(1990)95.}

\nref\rRoan{S.-S.~Roan: {\it Int.\,J.\,Math.}\,{\bf1}(1990)211,
     \I{\bf2}(1991)439.}

\nref\rGP{B.R.~Greene and M.R.~Plesser: \NP{B338}(1990).}

\nref\rKoszul{M.G.~Eastwood and T.~H\"ubsch: \CMP{132}(1990)383\semi
      P.~Berglund, T.~H\"ubsch and L.~Parkes: The Complete Matter
      Sector in a Three-Generation Compactification, \CMP{}(in
      press).}

\nref\rAdP{P.~Green and  T.~H\"ubsch: \CMP{115}(1988)231.}

\nref\rtex{P.~Candelas, A.M.~Dale, C.A.~L\"utken and R.~Schimmrigk:
     \NP{B298}(1988)493.}

\nref\rChiRi{C.~Vafa and N.~Warner: \PL{218B}(1989)51\semi
      W.~Lerche, C.~Vafa and N.~Warner: \NP{B324}(1989)427.}

\nref\rNoCon{T.~H\"ubsch: \CQG{8}(1991)L31.}

\nref\rGEPNER{D.~Gepner: \CMP{142}(1991)433-492}

\nref\rGRY{B.~Greene, S.-S.~Roan and S.-T.~Yau: Geometric
      Singularities and Spectra of Landau-Ginzburg Models.
     \Pre{Cornell} CLNS-91-1045 (1991).}

\nref\rSteenbrink{J.~Steenbrink: {\it Compos.\ Math.
     \bf34}(1977)211--223\semi see also
      S.~Cecotti: \IJMP{A6}(1991)1749.}

\nref\rCYHS{S.J.~Gates and T.~H\"ubsch: \PL{226}(1989)100,
     \NP{B343}(1990)741\semi see also
      J.I.~Latorre and C.A.~L\"utken: \PL{222B}(1989)55.}

\nref\rMarjD{T.~H\"ubsch: \MPL{A6}(1991)1553.}

\nref\rorbi{L.~Dixon, J.A.~Harvey, C.~Vafa and E.~Witten:
     \NP{B261}(1985)678, \I{B274}(1986)285\semi
      M.~Mueller and E.~Witten: \PL{182B}(1986)28\semi
      K.S.~Narain, M.H.~Sarmadi and C.~Vafa: \NP{279}(1987)369\semi
      L.E.~Iba\~nez, J.~Mas, H.P.~Nilles and F.~Quevedo:
     \NP{B301}(1988)157.}


\newsec{Introduction} 

Much work has been expended over the last few years on understanding
the geometrical content and interpretation of conformal field theories,
motivated largely  by the equivalence of the latter structure
with static classical string vacua. This has yielded a satisfying and
to some extent unexpectedly close connection between geometrical and
conformal field theoretical  constructs.
In the present paper we examine one perhaps somewhat technical
aspect of this fruitful correspondence.

In particular, we examine the physical manifestation of a well known
mathematical possibility initially studied in the context of string
compactifications in Ref.~\rPDM\,: situations exist for which there
are obstructions to faithfully representing all elements of the
$(2,1)$-cohomology group of a \CY\ manifold via polynomial
deformations. If such a Calabi-Yau manifold has a Landau-Ginzburg
interpretation this means that not all of the chiral primary fields
on the manifold are to be found in the (untwisted) chiral ring ${\cal
R} \subset P[X_i]/\Im$ where $P[X_i]$ is the ring of polynomials in
$X_i$ and $\Im$ is the ideal generated by the gradients of the
Landau-Ginzburg superpotential.  This is no surprise as we know that
such Calabi-Yau manifolds are related to {\it orbifolds} of
Landau-Ginzburg theories~\refs{\rDoron,\rGVW,\rMartinec} and hence
the conformal theory has twisted sectors in addition to the Jacobian
ring mentioned above. On general grounds, the correspondence between
Calabi-Yau manifolds and Landau-Ginzburg orbifolds assures us that,
in the end, these twisted sectors will account for the missing
cohomology.  This is somewhat reminiscent of what happens in the
study of singular Calabi-Yau spaces such as toroidal orbifolds in
which `blow-up modes' in twisted sectors of the conformal field
theory account for the additional cohomology associated with the
resolution of singularities.  In the present case, however, we are
dealing with twisted cohomology arising from cohomological
obstructions in the context of {\it smooth} manifolds.

Our intent here is to go beyond the by now universally accepted
belief that the mathematical and physical analyses converge to the
same answer, and determine to what extent the detailed structure of
graded twisted sectors in conformal theories finds expression in the
mathematical machinery of exact and spectral sequences.  We find a
remarkably close parallel between the `mathematical' and `physical'
descriptions which gives rise to a canonical map between the two.  In
fact, a natural and isomorphic ring structure may be imposed on the
elements in each formulation \foot{The relation between this ring
structure and the one given by the underlying conformal field theory
is as yet unknown, leaving the physical relevance of this isomorphism
unclear for now.}.  Furthermore, our Koszul complex calculations are
seen to provide the direct mathematical version of the analysis of
Intrilligator and Vafa~\rLGO, who used spectral flow to find explicit
representatives for the modes in \LGO\ models.  In essence, then, we
gain new insight into the mathematical significance of twisted
sectors in conformal field theory. Simultaneously, the success of our
comparison shows that a careful analysis of the point-field limit
continues to reveal surprisingly many details of the stringy theory.
We also briefly emphasize the {\it presentation} dependence of
whether a particular mode arises from a twisted or an untwisted
sector. Namely, a given abstract Calabi-Yau manifold can be realized
as embedded in different ambient projective spaces, being cut out by
different sets of defining equations.  Although the manifolds being
described are isomorphic, twisted modes in one presentation can arise
as untwisted modes in the other, and vice versa.  We explicitly
illustrate this phenomenon.

In particular, we show~: {\bf1.}~the \LGO\ analysis copes
gracefully with the situations in which the na\"\ii ve polynomial
deformations are ineffective~\rPDM\,: all extra states and
reparametrizations show up in the twisted sectors; {\bf2.}~the
grading in the particle spectrum provided by the twisting in the
\LGO\ analysis is recovered in the geometric analysis and provides
analogous selection rules; {\bf3.}~the above provides a natural
correspondence between all elements of the $(c,c)$ and $(a,c)$ rings
obtained via the spectral flow arguments of Refs.~\rLGO\ and the
cohomology elements obtained as particular tensor fields in the
Koszul complex formalism.

One further point of interest has to do with the mathematical
approach~\rRoan, to understanding the mirror manifold constructions
of Ref.~\rGP. This approach is firmly rooted in having an explicit
understanding of the cohomology of a given example in terms of
polynomial representatives. The brief study presented here together
with Refs.~\rLGO\ indicates that numerous theories, in which na\"\ii
ve polynomial deformation methods fail, submit to a generalized
polynomial representation.  This may well be helpful in extending the
methods of Ref.~\rRoan\ beyond the limited class to which it
presently applies.

Rather than burying ourselves in the notational complexity of the
general case, we present a rather non-trivial model which exhibits
all of the essential features common to examples of this sort.  We
proceed as follows. In section~2, we present the charged matter and
moduli sector of a rather non-trivial model from both the classical
geometry and the \LGO\ point of view. To save some space, we refer
the reader to Refs.~\rLGO\ for the \LGO\ analysis and to
Refs.~\rKoszul\ for the computational techniques of
classical geometry.  In section~3, we describe the geometry of the
\LGO, aiming to explain the high degree of agreement which we find.
Finally, in section~4, we discuss the present results and comment on
some related issues.

\newsec{A Warped Model} 
As a representative example, consider the \CY\ manifold
\eqn\edefeqs{
  {\cal M} \in \K{3\cr2\cr1\cr}{3&1&0\cr 0&2&1\cr 0&0&2\cr}~~ :
 ~~\left\{
       \eqalignthree{
   f(x)   &= f_{abc}\, x^a\, x^b\, x^c
           &= ~\sum_{a=0}^3 (x^a)^3       &= ~0~,  \cr
   g(x,y) &= g_{a\,\a\b}\, x^a\, y^\a\, y^\b
           &= ~\sum_{\a=0}^2 x^\a(y^\a)^2 &= ~0~,  \cr
   h(y,z) &= h_{\a\,rs}\, y^\a\, z^r\, z^s
           &= ~\sum_{r=0}^1 y^r(z^r)^2    &= ~0~,  \cr}\right.}
where $x,y,z$ are homogeneous coordinates on the space $\CP3 \times
\CP2 \times \CP1$; in the {\it configuration matrix} notation above,
it is represented by the left most `bra'-like column listing the
dimensions.  The `ket'-like degree matrix to the right of `$\|$'
represents the constraints~: the first column for $f(x)$, a cubic in
\CP3, the second column for $g(x,y)$---of bi-degree (1,2) in $\CP3
\times \CP2$ and the third column for $h(y,z)$---of bi-degree (1,2)
in $\CP2 \times \CP1$. Clearly, the coefficient tensor $f_{abc}$ is
totally symmetric, $g_{a\,\b\g} = g_{a\,\g\b}$ and $h_{\a\,rs} =
h_{\a\,sr}$. From Ref.~\rAdP, we have that $\EU=-48$, $b_{1,1}=9$ and
so $b_{2,1}=33$.

Polynomial deformations are easily checked to parametrize only 24 of
the {\bf27}'s and the three K\"ahler forms of \CP3, \CP2 and \CP1
span only one third of the {\bf27}\con's. We therefore expect some
higher forms to represent both the nine missing {\bf27}'s and the six
missing {\bf27}\con's.

\subsec{Classical geometry}
As mentioned above, the cohomology of the \CY\ manifolds is obtained
using the Koszul complex and the Bott-Borel-Weil theorem, which
relies on the fact that
\eqn\eCPn{
     \CP{n}~ \approx ~{U(n{+}1) \over U(1) \times U(n)}~.}
It follows that all homogeneous bundles over \CP{n} are specified by
$U(1) \times U(n)$ representations while all cohomology on \CP{n} is
specified by $U(n{+}1)$ representations. This then implies that
harmonic forms on $\cal M$, and thus the {\bf27}'s and {\bf27}\con's,
correspond to certain {\it components} of $U(4) \times U(3) \times
U(2)$ tensors, as listed below. For details of this technique
see Refs.~\rKoszul. Notation~: we use indices
$a,b,\ldots$ for $U(4)$, $\a,\b,\ldots$ for $U(3)$ and $r,s,\ldots$
for $U(2)$; so for example $\f^{(ab)}$ is a totally symmetric rank-2
contra-variant $U(4)$-tensor; note that contra- and co-variant (upper
and lower) indices must be distinguished.

Upon the straightforward calculation (following Refs.~\rKoszul), we
obtain that the {\bf27}\con's are parametrized by $J_x$, $J_y$,
$J_z$, (the pullbacks of) the three K\"ahler classes from \CP3, \CP2
and \CP1 and the six tensor components
\eqn\eextra{
     \{\> \ve^{0123} \f^{(ab)}~ : ~ a\ne b~,~ a,b=0,\3,3 \>\}_6~.
}

As for the {\bf27}'s, the Koszul complex computation represents
\hm{1}{\TM} as
\eqn\eXXX{
    \matrix{
 \big[\,\{\, \f_{(abc)} \,\}/
         \{\, f_{d(ab}\, \l_{c)}{}^d \,\}\,\big]~~
 \oplus
 ~~\big[\,\{\, \vf_{a(\b\g)} \,\}/
           \{\, g_{d\,\b\g}\, \l_a{}^d \oplus
                g_{a\,\d(\b}\, \l_{\g)}{}^\d \,\}\,\big]  \cr
 \oplus ~~\big[\,\{\, \j_{\a(rs)} \,\}/
                  \{\, h_{\d\,rs}\, \l_\a{}^\d \oplus
                       h_{\a\,t(r}\, \l_{s)}{}^t \,\}\,\big]~~
 \oplus ~~\big[\,\{\, \ve^{01}\,\vq_{a\,\b} \,\}/
                  \{\, \ve^{01}\, g_{a\,\b\g}\, \l^\g
                                            \,\} \,\big]~.\cr}
}
The variables $\l_a{}^b$, $\l_\a{}^\b$, $\l_r{}^s$ and $\l^\a$ are
`reparametrization' degrees of freedom which can be used to `gauge
away' 16, 9, 4 and 3 components, respectively. Given our choice of
$f_{abc}$, $g_{a\,\b\g}$ and $h_{\a\,rs}$ in~\edefeqs, a
convenient basis is provided by the following 33 tensor components
\eqn\etensors{
    \eqalign{
 H^1({\cal M},{\cal T_M})
 &\sim \{\, \f_{(abc)}~ : ~a\ne b\ne c \,\}_4~~
   \oplus ~~\{\, \vf_{a(\b\g)}~ : ~\b\ne a\ne\g \,\}_{15}~,  \cr
 & ~\oplus ~~\{\, \j_{\a(rs)}~ : ~r\ne\a\ne s \,\}_5~~
     \oplus ~~\{\, \ve^{01}\,\vq_{a\,\b}~ : ~a\ne\b \,\}_9~. \cr}
        }
Contracting with $x$'s, $y$'s and $z$'s, the above
representatives are dual to the monomials
\eqn\eXXXV{
    \matrix{
 \{\, (x^a\,x^b\,x^c)~ : ~a\ne b\ne c\ne a \,\}_4~, &
 \{\, (x^a\, y^\b\,y^\g)~ :    ~a\ne\b,\g \,\}_{15}~,     \cr
 \{\, (y^\a\, z^r\,z^s)~ : ~r\ne\a\ne s \,\}_5~, &
 \{\, (x\rd^3x)(y\rd^2y)(x^a\, y^\b)~ : ~a\ne\b \,\}_9~.  \cr}
}
As the complementary factor in the holomorphic part of the total
volume form, $\ve^{01} \sim (z\rd z)$ is dual to
$(x\rd^3x)(y\rd^2y)$.

The first $24$ representatives are the usual polynomial
deformations. The last nine and also the six ones in~\eextra,
however, arrive as elements of higher cohomology groups.

\subsec{The Landau-Ginzburg orbifold}
We now consider the superpotential
\eqn\esuperW{
     W~ \define ~\sum_{r=0}^1
                 (X_r{}^3 + X_r\, Y_r{}^2 + Y_r\, Z_r{}^2)~ +
                ~(X_2{}^3 + X_2\, Y_2{}^2)~ + ~X_3{}^3~,
}
corresponding to the system~\edefeqs. We check that the
origin in the field-space is the only critical point and that the
central charge is $c=9$, since $q(X_a) = q(Y_\a) = q(Z_r) = \inv3$.
Also,
\eqn\eXXX{
    \eqalign{
     \det\big[ \vd^2W \big]
 &= 2^9\cdot 3\cdot \prod_{r=0}^1
                 (3X_r{}^2 Y_r - Y_r^3 - 3X_r Z_r{}^2)
                 (3X_2{}^2 - Y_2^2) (X_3)~,      \cr
 &\equiv 2^{14}\cdot 3^4\cdot X_0{}^2 Y_0\, X_1{}^2 Y_1\,
             X_2{}^2\, X_3~ \pmod{\Im[\vd W]}~,  \cr}
}
the non-vanishing of which guarantees the existence of the highest
weight field in the superconformal field theory into which the
\LGO\ renormalizes.\ping

We now analyze the spectrum of light charged matter fields
following Refs.~\rLGO. We will need the general formula for the
charges of the Ramond vacuum in the $\ell^{th}$ twisted sector~:
\eqn\eRamondCh{
  {J_0\atop\Jb_0} \Big|0\Big\rangle^{(\ell)}_R~ = \bigg\{\,
   \pm\Big[\sum\limits_{\Q_i(\ell) \not\in \ZZ}
               (\Q_i(\ell) - [\Q_i(\ell)] - {1\over2})\Big]~ +
     ~\Big[\sum\limits_{\Q_i(\ell) \in \ZZ}
               (q_i - {1\over2})\Big] \,\bigg\}
                                  \Big|0\Big\rangle^{(\ell)}_R~,
}
where $\Q_i(\ell)$ is the twisting angle of the $i^{th}$ field in the
$\ell^{th}$ sector. The matching $\big|0\big\rangle^{(\ell)}_{(c,c)}$
and $\big|0\big\rangle^{(\ell+1)}_{(a,c)}$ vacua are obtained by
spectral flows ${\cal U}_{(1/2,1/2)}$ and ${\cal U}_{(-1/2,1/2)}$, of
charges $({3\over2},{3\over2})$ and $(-{3\over2},{3\over2})$,
respectively; note the $\ell\to\ell+1$ shift in the ${\rm
Ramond}\to(a,c)$ flow.

\subsec{The warp}
As in Ref.~\rGVW, we can find a holomorphic transformation with
constant Jacobian to identify the necessary $\ZZ_n$-twist. Since
all fields scale with $q=\inv3$, a $\ZZ_3$-twisting would seem
appropriate. In fact however, one must twist by a $\ZZ_{12}$
which can be found by a careful analysis as was done in Ref.~\rGVW; in
section~3, we will give a coordinate-independent and general
definition of this twist group. Suffice it here that we need the
\eqn\eXXX{
     (      X_a,         Y_\a,    Z_r )~ \cong
    ~( \w^4 X_a, \w^{10} Y_\a, \w Z_r )~,\qquad  \w^{12}=1
}
$\ZZ_{12}$ identification for the Landau-Ginzburg model to
correspond to the \CY\ manifold $\cal M$. To distinguish from the
scaling charges $q_x=q_y=q_z=\inv3$, we will write $\q_x=\inv3$,
$\q_y={5\over6}$ and $\q_z=\inv{12}$. We also record
$n_x=n_y=n_z=1$ and $d=3$ and that now $\Q_a(\ell)=\ell q_x$, but
$\Q_\a(\ell) = \ell\q_y \ne\ell q_y$ and $\Q_r(\ell) = \ell\q_z
\ne\ell q_z$.

\bigskip
\noindent{\bf The untwisted sector}:\ \
 From Eq.~\eRamondCh, we have that the left-moving and
right-moving $U(1)$ charges of the untwisted Ramond vacuum are $J_0
\big|0\big\rangle^{(0)}_R = \Jb_0 \big|0\big\rangle^{(0)}_R = -{3\over2}
\big|0\big\rangle^{(0)}_R$. Spectral-flowing to the N.S.\ sector, we get a
charge-(0,0) $(c,c)$ vacuum. All fields have classical solutions and
we list explicit representatives from
\eqn\eXXX{
 \bigotimes_{r=0}^1
             \{\, \Ione,\, X_r,\, Y_r,\, Z_r,\, X_r{}^2,\,
                           X_r Y_r,\, X_r Z_r,\, X_r^2 Y_r \,\}~
    \otimes ~\{\, \Ione,\, X_2,\, Y_2,\, X_2{}^2 \,\}~
    \otimes ~\{\, \Ione,\, X_3 \,\}~.
}
To obtain states of $(J_0,\Jb_0)$-charge (1,1) from the untwisted
charge-(0,0) vacuum, we need elements from $(X \oplus Y \oplus
Z)^{\otimes3}$. Requiring also integrality of the total $\Q$-twist
then leaves
\eqn\eXXX{
     \matrix{
 \{\, (X_a\,X_b\,X_c) : a\ne b\ne c\ne a \,\}_4             \cr
  \oplus ~\{\, (X_a\, Y_\b{}^2)~ :    ~a\ne\b \,\}_9~~
  \oplus~~\{\, (X_a\,Y_\b\,Y_\g)~: ~a\ne b\ne c\ne a \,\}_6 \cr
 ~\oplus ~\{\, (Z_r{}^2\,Y_s)~ :    ~r\ne s \,\}_4~~
  \oplus~~\{\, (Z_0 Z_1 Y_2) \,\}_1~,                       \cr}
}
These 24 representatives match {\it state-by-state} the first 24
representatives from~\eXXXV, the polynomial deformations.

\bigskip
\noindent{\bf The twisted sectors}:\ \
For $\ell=1,2,4,5,7,8,10,11$, no field is left invariant under the
$\Q$-twist and there are only the twisted vacua in these sectors.
The $\ell=2,7,11$ yield charge-($-1,1$) $(a,c)$-states which we
identify with (linear combinations of) $J_x$, $J_y$ and $J_z$. For
$\ell=3$, the $\Jb_0$-charge of the $(c,c)$- and $(a,c)$-vacuum is
${4\over3}$ and is too high to produce marginal $(c,c)$- or
$(a,c)$-states.

In the sector $\ell=6$, the $X_a$'s and the $Y_\a$'s are left
invariant and so have classical solutions. The $J_0$- and
$\Jb_0$-charges of the Ramond vacuum are both $-{7\over6}$,
yielding a $(J_0,\Jb_0)$ charge-$(\inv3,\inv3)$ vacuum. Clearly, we
need fields from $(X \oplus Y)^{\otimes2}$ to create marginal
$(c,c)$-states. Using Eq.~\eRamondCh\ replacing however $q_i
\to \q_i$, we find both the left and the right $\ZZ_{12}$-twist of
the $\ell=6$ Ramond vacuum to be \inv3, giving rise to $\Q
\big|0\big\rangle^{(6)}_{(c,c)} = \bar{\Q}
\big|0\big\rangle^{(6)}_{(c,c)} = {11\over6}
\big|0\big\rangle^{(6)}_{(c,c)}$. For marginal $(c,c)$-states of
integral $\ZZ_{12}$-twist, we can therefore use only
\eqn\eXXX{
     \{\, (X_a Y_\a)~ : ~a\ne\a \,\}_9~.
}
These nine states are in precise state-by-state correspondence to the nine
additional contributions in~\eXXXV. Furthermore,
$\Q((x\rd^3x)(y\rd^2y)) = {23\over6}$, which is indeed congruent,
modulo integers, to the $\Q$-charge of the $\ell=6$ twisted
$(c,c)$-vacuum. Thus, $\ve^{01}\big|0\big\rangle^{(0)}_{(c,c)} \sim
\big|0\big\rangle^{(6)}_{(c,c)}$, that is, $\ve^{01}$ acts as the
$\ell=6$ twist-field.

Finally, the Ramond vacuum in the $\ell=9$ twisted sector has
$(J_0,\Jb_0)$-charges $(-\inv6,-{7\over6})$, yielding a
charge-$(-{5\over3},\inv3)$ $(a,c)$-vacuum. Again, we need
quadratic fields to produce marginal $(a,c)$-states. The
$(\Q,\bar{\Q})$-twists now coincide with the $(J_0,\Jb_0)$-charges
and we easily list
\eqn\eXXX{
     \{\, (X_a X_b)~ : ~a\ne b \,\}_6~.
}
The six additional tensors in~\eextra\ all feature
$\ve^{0123} \sim (x\rd^3x)$. Now, indeed, $\Q(x\rd^3x) = {4\over3}
\id \inv3 \pmod{\ZZ}$, in accord with our identification~:
$\ve^{0123}\big|0\big\rangle^{(0)}_{(c,c)} \sim
\big|0\big\rangle^{(9)}_{(c,c)}$, that is, $\ve^{0123}$ acts as the
$\ell=9$ twist-field.

The upshot of this discussion is that the $\ve$ factors {\it
formally} act as twist fields, creating the twisted vacua from the
untwisted one. Their twist charge, computed from their dual monomial
representation, precisely equals the charge of the associated twisted
vacuum. Thus, we have a direct correspondence between these conformal
field theoretic twisted sectors and the higher cohomology elements
which account for the obstructions to polynomial deformation
analysis. Determining the precise details of this relation between
the $\ve$ factors and the twist fields of superconformal field theory
appears worthwhile, but we defer this to future work. Suffice it here
to note that the `mathematical' and `physical' analyses, although
vastly different in origin, present precisely isomorphic data.

\subsec{Ineffective splitting}
The reader might have been lulled into believing that the untwisted
sector of a \LGO\ is synonymous to polynomial deformations in
classical geometry, whereas the twisted $(c,c)$-states give rise to
``higher'' deformations of the complex structure and of course, to the
K\"ahler variations in the $(a,c)$-sector. Rather than appeal to
general arguments why this need not be the case, we suggest
analyzing the superpotential
\eqn\eXXX{
     W~ \define ~\sum_{r=0}^1
    (X_r{}^3 + X_r\, Y_r{}^2 + Y_r\, V_r + V_r Z_r{}^2)~ +
   ~(X_2{}^3 + X_2\, Y_2{}^2)~ + ~X_3{}^3~,
}
which corresponds to
\eqn\eXXX{
  {\cal Y} \in \K{3\cr2\cr1\cr1\cr}
               {3&1&0&0\cr 0&2&1&0\cr 0&0&1&1\cr 0&0&0&2\cr}~,
 \qquad\eqalignthree{
  q_x={1\over3}~,~~ & ~~ q_y={1\over3}~,~~ &
                  ~~ q_z={1\over6}~,~~ & ~~ q_v={2\over3}~,  \cr
 \q_x={1\over3}~,~~ & ~~\q_y={5\over6}~,~~  &
                  ~~\q_z={5\over12}~,~~ & ~~\q_v={1\over6}~. \cr}
}
The erudite reader may have noticed that this is an ``ineffective
split''~\rtex\ of the previous model, $\cal M$, where the
`ineffective split' means that although the ambient space has been
enlarged and the set of defining equations altered, nevertheless
${\cal Y} = {\cal M}$.

The matter spectrum now contains the untwisted charge-$(1,1)$ states
(with suitable restrictions on the indices)
\eqn\eXXXII{
     \{ X_a X_b X_c \}_4~ \oplus ~\{ X_a Y_\b Y_\g \}_{15}~ \oplus
    ~\{ Y_\a V_r \}_4~ \oplus ~\{ X_a Y_\b Z_0 Z_1 \}_9~ ,
}
a single $\ell=6$ charge-$(1,1)$ state, $\{ Y_2 \}_1$, and nine
twisted charge-$(-1,1)$ states~:
\eqn\eXXX{
     \{ X_a X_b \}_6^{{\rm from}~\ell=9}~ \oplus
     \{ \hbox{twisted vacua from } \ell=7,8,11 \}_3~.
}
Of these, the first 23 in~\eXXXII\ are realized as polynomial
deformations. The remaining nine untwisted and the one twisted
charge-$(1,1)$ modulus cannot be realized as a deformation of the
embedding ${\cal Y} \into \CP3 \times \CP2 \times \CP1_z \times
\CP1_v$ although, of course, all 33 charge-$(1,1)$ moduli correspond
to independent deformations of the complex structure.

The Koszul complex computation again lists the last nine untwisted
charge-$(1,1)$ moduli as stemming from higher cohomology and they are
of the form
\eqn\eXXX{
    (\ve^{01}_{(Z)} \ve^{01}_{(V)} \vq_{a\b})
 ~~ \buildrel{*}\over{\sim} ~~
    (x\rd^3x)(y\rd^2y) x^a y^\b~.
}
In fact, comparison with the results for $\cal M$ shows that $(Z_0
Z_1)$ plays the r\^ole of a twist field. That is,
\eqn\eXXX{
    \left[\, X_a Y_\b Z_0 Z_1 \big| 0 \big\rangle^{(0)}_{(c,c)}~
      = ~X_a Y_\b \big| Z_0 Z_1 \big\rangle^{(0)}_{(c,c)}
            \,\right]_{{\rm in}~\cal Y}
   ~~ \sim ~~
    \left[\, X_a Y_\b \big|0 \big\rangle^{(6)}_{(c,c)}
           \,\right]_{{\rm in}~\cal M}~,
}
which agrees with the respective charges and
$\ZZ_{12}$-twists (modulo $\ZZ$, of course).

The main utility of this phenomenon lies in the possibility to
represent the twisted $(c,c)$-moduli of the \LGO\ $\cal M$ as untwisted
moduli of the ``ineffectively split'' variant thereof, $\cal Y$, at the
expense of one untwisted $(c,c)$-modulus of $\cal M$ which becomes
twisted in $\cal Y$. Since untwisted fields can be added to the
superpotential to deform the complex structure, we may study such
deformations by examining the effect of
\eqn\eXXX{
     W~ \to ~W + t^{abc}   (X_a X_b X_c)
               + t^{a\b\g} (X_a Y_\b Y_\g)
               + t^{\a r}  (Y_a V_r)
               + t^{a\b}   (X_a Y_\b Z_0 Z_1)~,
}
even though the last nine terms do not represent conventional
polynomial deformations.

\subsec{The ideals match ideally}
The diligent reader will have noticed a possible embarrassment in
identifying the classical model based on the \CY\
manifold~\edefeqs\ and the \LGO\ with the
superpotential~\esuperW. In both cases, the states are
described as certain polynomials modulo some ideal. In general, the
generators  of the ideal are rather different in the two approaches
as soon as there is more than one constraint and this calls for a
closer look.

On one hand, the states in the \LGO\ analysis are polynomials in
those superfields which satisfy the twist conditions for the
particular twisted sector, taken however modulo the ideal generated
by the gradients of the superpotential\foot{More properly, the ideal
is generated by the equations of motion. With a kinetic term
$K(X,\bar{X})$, this implies $\vd_a W = \bar{D}^2[\vd_a
K(X,\bar{X})]$, which involves both left- and right-handed
$\bar{D}$. So, if $\vd W$ appears in a correlation function where
only chiral and/or twisted-chiral superfields occur (which are
annihilated by one of the two $\bar{D}$'s), we can `pull a
derivative over' to annihilate the entire correlator.}.

On the other hand, the quotients in~\etensors\ involve the
ideal generated by $\vd f(x)$, $\vd g(x,y)$ and $\vd h(y,z)$. To
see this, simply note that
\eqna\eXXX
$$
\eqalignno{
  f_{d(ab} \l_{c)}{}^d\cdot x^a\, x^b\, x^c
 &= \l^a(x)\,\vd_a f(x)~,\qquad\hskip3.7mm
           \l^a(x) = \l_b{}^a\, x^b~,               & \eXXX a \cr
  g_{d\,\b\g} \l_{a}{}^d\cdot x^a\, y^\b\, y^\g
 &= \l^a(x)\,\vd_a g(x,y)~,                         & \eXXX b \cr
  g_{a\,\d(\b} \l_{\g)}{}^\d\cdot x^a\, y^\b\, y^\g
 &= \l^\a(y)\,\vd_\a g(x,y)~,\qquad
           \l^\a(y) = \l_\b{}^\a\, y^\b~,           & \eXXX c \cr
  h_{\b\,rs} \l_{\a}{}^\b\cdot y^\a\, z^r\, z^s
 &= \l^\a(y)\,\vd_\a h(y,z)~,                       & \eXXX d \cr
  h_{\a\,t(r} \l_{s)}{}^t\cdot y^\a\, z^r\, z^s
 &= \l^r(z)\,\vd_r h(y,z)~,\qquad
     \l^r(z) = \l_s{}^r\, z^s~.                     & \eXXX e \cr}
$$
Clearly, $\l^a(x)$, $\l^\a(y)$ and $\l^r(z)$ span linear
reparametrizations. This would seem to imply that all  fields
proportional to $\vd f(x)$, $\vd g(x,y)$ or $\vd h(y,z)$ become
`gauged away', that is, that fields are taken modulo the ideal
$\Im[\vd_a f, \vd_a g, \vd_\a g, \vd_\a h, \vd_r h]$.

The full ideals $\Im[\vd_a W,\vd_\a W,\vd_r W]$ and
$\Im[\vd_af, \vd_a g, \vd_\a g, \vd_\a h, \vd_r h]$
can be listed easily and their difference (on states of unit
scaling charge) is generated by
\eqn\eXXX{
     [\vd_a f(x) - \vd_a g(x,y)]
      \qquad{\rm and}\qquad
     [\vd_\a g(x,y) - \vd_\a h(y,z)]~.
}
So these two ideals indeed differ substantially while, in contrast,
the list of massless states had been obtained completely identical.

The resolution of this lies in the simple fact that there are only
$16{+}9{+}4$ degrees of freedom for such `gauge choices'
in~\etensors\ and that there were distinct but equivalent `gauge'
choices. This ambiguity was taken care of in the Koszul complex by
noting that the map $\bigoplus_{a=f,g,h} {\cal E}_a\con \to {\cal
O}_{\cal W}$ has `gauge degrees of freedom'---a kernel represented by
$\{ {\cal E}_a\con \wedge {\cal E}_b\con\}_{a,b=f,g,h}$. Finally, the
map $\wedge^2{\cal E}_\star\con \to \bigoplus_a{\cal E}_a\con$ itself
has a kernel, represented now by $\bigwedge_{a=f,g,h}{\cal E}_a\con$.

In other words, not all elements of the ideal generated by $\vd f
\oplus \vd g \oplus \vd h$ are effective---only those which are
non-trivial modulo $[\vd_a f-\vd_a g]$ and $[\vd_\a g-\vd_\a h]$. But
precisely this redundancy is the difference between the na\"\ii ve
`classical' ideal and $\Im[\vd W]$~\foot{There is no redundancy
proportional to say $[\vd_r g-\vd_r h]$ since $\vd_r g = 0$ and
$[\vd_r h]$ is a generator in both classical and quantum ideal.  That
is, e.g. $[\vd_a f-\vd_a g]\cong0$ reduces the ideal from $\{[\vd_a
f],[\vd_a g]\}$ to $[\vd_a f + \vd_a g]$ while, e.g. $[\vd_r g-\vd_r
h]\cong0$ says nothing new since $\vd_r g = 0$ and $[\vd_r h]\cong0$
anyway.}.  So, the true classical ideal, as implemented by the Koszul
complex, {\it precisely equals} the \LGO\ ideal, whence the spectrum
of states naturally becomes equal. Thus, not only do the massless
fields as found by classical geometry and by the \LGO\ techniques
match one--to--one, a natural and isomorphic ring structure may be
imposed on each. As mentioned earlier, an interesting question is to
work out the relationship of such a ring structure with the one
determined by conformal field theory.  This has only been done so far
in the untwisted $(c,c)$ sector in which they are identical.

\newsec{Why is Quantum Geometry so Classical?} 
The analysis so far has revealed a striking similarity between the
classical description of \hm{\star}{\TM} and the quantum \LGO\
description of the corresponding fields. We now discuss this a
little further and provide an explanation of it.

\subsec{Landau-Ginzburg orbifolds for complete intersections}
Although we have demonstrated this only in an explicit example, it is
generally true that both \hm{1}{\TM} and \hm{2}{\TM} of a complete
intersection manifold $\cal M$ in products of (weighted) homogeneous
spaces admit a parametrization of the form
\eqn\eTMcoho{
     \bigoplus_{k}
      \big\{\> P_k[x_i]\, \big/ \,\Im_k \>\big\}^{(k)}~.
}
For each $k$, $P_k[x_i]$ is a polynomial ring in certain suitable
variables $x_i$, $\Im_k$ is an ideal and $k$ labels a grading. This
is so for example in the Koszul complex computation of Refs.~\rKoszul.

The general analysis of N=2 supersymmetric conformal field theories
shows that the moduli fields also span such structures, the so-called
chiral and twisted-chiral rings~\rChiRi, which in cases of \LGO s,
feature a grading~\rLGO\ similar to the one found above. A crucial
point in the \LGO\ analysis is however that the ideals $\Im_k$ are in
fact all parts of one ideal generated by gradients of a {\it single}
function, the superpotential. On the other hand, given a \CY\
complete intersection, the Koszul complex computation yields the
analogous ideal(s) rather differently. We are therefore led to
conclude the following general fact~:
\bigskip

\vbox{\narrower
\it\noindent
Let $\cal M$ be a \CY\ space for which the \hm{1}{\TM} and \hm{2}{\TM}
cohomology groups are parametrized as in~\eTMcoho. Then

\item{1.} $\cal M$ has a \LGO\ description precisely if the
generators of $\Im_k$ can be integrated to a single function $W$,
critical only at the origin;

\item{2.} $W$ is the superpotential of the corresponding \LGO.}
\nobreak\vglue1pt
(A function is critical where its inverse is singular.)

Note that the variables $x_i$ in~\eTMcoho\ become coordinates
in the corresponding \LGO\ even if they were merely formal variables
parametrizing the cohomology. In fact, the above statement
generalizes the results of Ref.~\rGVW\ in the sense that a \LGO\ is
assigned to a \CY\ space $\cal M$ based on the cohomology
parametrization~\eTMcoho---independently of any particular
embedding of $\cal M$ and even if no embedding is known. Through the
geometric description below, this may provide (perhaps a different
realization) of the \CY\ manifold that we started from.

Ref.~\rGVW\ lists several necessary conditions for the existence of a
\LGO\ corresponding to a given complete intersection in products of
complex projective spaces and provides a construction of the
appropriate superpotential. Unfortunately, however, explicit analysis
of a special class of complete intersections (even just in products
of complex projective spaces) shows that these conditions are not
sufficient~\rNoCon\ and that the construction of Ref.~\rGVW\ may
yield degenerate \LGO s. We have now shown that the problem lies in
integrating the generators of the ideal(s) $\Im_k$ to a single
function\foot{Related results from the superconformal field theory
vantage point may be found in Ref.~\rGEPNER.} and we hope to return
to this and provide at least a partial solution in a future account.

\subsec{The geometry of Landau-Ginzburg orbifolds}
We now analyze the \LGO\ construction in terms of classical geometry
in order to explain the structural and very detailed identification
presented above.

Rather crucially, the kinetic terms are ignorable in the analysis of
Refs.~\refs{\rGVW, \rChiRi} and the scalar component-fields are maps from
the world sheet into an {\it affine space}. For example, in
Eq.~\esuperW, the $(X_a,Y_\b,Z_r)$ map into $\IC^9 \approx \IC^4
\times \IC^3 \times \IC^2$. The I\,R regime of the Landau-Ginzburg
model concentrates on the hypersurface $W=0$, which is a {\it cone}
in $\IC^9$.

The classical description instead relates to the \CY\ manifold ${\cal M}
\subset \CP3 \times \CP2 \times \CP1$; of course, $\CP{n} \approx
\{(\IC^{n+1}-0) / \IC^*\}$, where $\IC^*$ denotes homogeneous scaling
by $\l\ne0$. So, $\cal M$ corresponds not to the Landau-Ginzburg cone
in $\IC^9$, but to the quotient thereof by the $\IC^*_x \times
\IC^*_y \times \IC^*_z$-action. However, $W$ is invariant only with
respect to a finite subgroup $\Q \subset \IC^*_x \times \IC^*_y
\times \IC^*_z$, the action of which is found as the solution to
$\l_x{}^3 = \l_x\,\l_y{}^2 = \l_y\, \l_z{}^2 = 1$.

This then finally defines the correct twist group for the \LGO~: $\Q$
is simply {\it~the maximal finite subgroup of the rescaling group
$\prod_i \IC^*_i$ which leaves the superpotential strictly
invariant}\foot{Several actions of $\ZZ_{12}$ on $W$~\esuperW\
emerge, corresponding to different prime roots in $\ZZ_{12}$; they
are of course equivalent and yield the same results.}. The so defined
twist group for the \LGO\ is {\it exactly the same} as the one
defined in Ref.~\rGVW. There, one performs a change of variables which
turns redundant coordinate fields into Lagrange multipliers,
requiring that the Jacobian be field-independent; in fact, the
Jacobian {\it equals} $|\Q|$, the order of the twist group,
indicating that the field-space has been orbifolded into $|\Q|$
copies of a fundamental domain. Finally, the same twist group can
also be defined by requiring that the field-space of the \LGO\ admit
a global (path-)integration measure~\rGRY.\ping

It may appear unexpected that $\Q$ is required to leave the
superpotential, defining the cone in $\IC^9$, {\it strictly} rather
than just {\it projectively} invariant.  This follows from two
different, although perhaps not unrelated points of view.

On one hand, a standard trick~\rSteenbrink\ for computing the
cohomology of hypersurfaces $W(x_i)=0$ in an affine embedding space
is to~: (1)~consider level sets $\{ W(x_i) = \m,~ \m\in\IC\}$ for
$\m$ a new variable, then (2)~reinterpret this as a hypersurface
$W(x_i)-\m=0$ in $\IP(x_i,\m)$. In the patch $(\m\ne0)$, rescale
$\m\to1$ obtaining $W(x_i)=1$, while at the ``infinity'' of that
patch $(\m=0)$ the original $W(x_i)=0$ is recovered. Since results in
different patches are related, we must preserve $W(x_i)=1$, hence the
requirement of {\it strict} rather than {\it projective} invariance
with respect to $\Q$ and our above definition of $\Q$ follows.

 From a field theory point of view, for every complete intersection
manifold $\cal M$, there is a constrained $\s$-model~\rCYHS. The
supersymmetric action is
\eqn\eXXX{
  S_{\cal M}~ \define ~\int_\S \rd^2\s
 \big\{\, {\cal L}_{\rm kin.} + {\cal L}_{\rm con.} \,\big\}~,\qquad
      \left\{
     \eqalign{
  {\cal L}_{\rm kin.} &\define {\ddd\int} \rd^4\vs
              \sum_{r=1}^m w^r K_r(X,\Xb)~,              \cr
  {\cal L}_{\rm con.} &\define {\ddd\int} \rd^2\vs
              \sum_{a=1}^K \L_a P^a(X)~~ + ~~{\rm h.c.}  \cr}\right.
}
$K_r(X,\Xb)$ is the scale-invariant K\"ahler potential for \CP{n_r}
and $w^r$ is its relative ``size''; the scale-invariance can be made
explicit by introducing a gauge superfield $V_r$ for each \CP{n_r}.
$P^a(X)$ is the $a^{th}$ constraint polynomial and $\L_a$ the
corresponding Lagrange multiplier superfield.  Path-integration over
the $\L_a$ enforces $P^a(X)=0$ and restricts $\prod_r\CP{n_r}$
to $\cal M$.

Without the constraints, the $\s$-model describes $\prod_r\CP{n_r}$,
is invariant under the $\prod_{r=1}^m \IC^*_r$ gauge symmetry and, as
usual in field theories, we must divide out the action of this group.
The constraints break the $\prod_{r=1}^m \IC^*_r$ gauge symmetry
{\it explicitly} to the maximal subgroup which leaves the full action
$S_{\cal M}$, thus each polynomial {\it strictly invariant}. As the
$P^a(X)$ are non-linear, $\Q$ is inevitably finite; it depends only
on the degrees of $P^a(X)$ and the weights of $X_i$. Being a remnant
of a gauge symmetry, $\Q$ must be ``modded out''.

In the gauge where $V_r=0$ and $\L_a=1$, $S_{\cal M}$ becomes the
standard Landau-Ginzburg action~\rMarjD. It is easy to see that
$\Q$ is precisely that symmetry of the Landau-Ginzburg action which
has been modded out in the approaches of Refs.~\rGVW\ and~\rGRY.

\subsec{Towards a universal understanding of twisted modes?}
There is one more point requiring clarification before we finish this
section, regarding the occurrence of twisted modes. In the above
analysis, twisted modes have shown up as related to certain
cohomological obstructions whereas in Coxeter orbifolds~\rorbi, they
occur corresponding to exceptional divisors in a blow-up. We here
suggest another possible viewpoint, in which twisted modes indeed
correspond to exceptional divisors. At present, this point of view is
easily adopted for simple models like the quintic hypersurfaces in
\CP4, but we feel that a more universal treatment may be developed in
time.

In the spectrum of the \LGO\ orbifold associated to the quintic in
\CP4, there are four twisted states, all in the $(a,c)$-sector and
one of which is marginal, i.e., has charge-$({-}1,1)$. This
``K\"ahler twisted state'' corresponds to the pull-back of the
K\"ahler form of \CP4 and so to the intersection of the quintic
hypersurface with a hyperplane in \CP4.  So far, this does not appear
to be the exceptional divisor of any blow-up.

However, recall that the dual of the hyperplane bundle,
${\cal O}({-}1)$ over \CP{n} is in fact the blow-up of $\IC^{n+1}$ at
the origin. The total space of ${\cal O}({-}1)$ over \CP{n} is
non-compact and has $\IC^1$-like fibres; \CP{n} itself is the
exceptional divisor created by blowing up. Moreover, away from the
origin, $\IC^{n+1}$ looks just like the total space of ${\cal O}({-}1)$
away from the \CP{n}. Next, the total space of ${\cal O}({-}k)$ looks
the same, except that the fibres at $x\in\CP{n}$ vary as a degree-$k$
rather than a linear polynomial in $x$ and are hence related to the
original ${\cal O}({-}1)$ fibres in a $k$--1 manner\,: only up to a
$\ZZ_k$-twist. Again, \CP{n} itself is the exceptional divisor which
replaced the origin.

Consider now a quintic hypersurface in \CP4, $\cal M$. The \LGO\
would involve the corresponding quintic in $\IC^5$, $M$. Blowing this
up at the origin, we obtain $\Tw{M}$, a quintic in the total space of
${\cal O}({-}1)$, which is also a quintic in ${\cal O}({-}5)$ upon a
$\ZZ_5$-twist and which is easily seen to be a fibre space with open
$\IC^1$-like fibres over $\cal M$. Therefore, the (co)homology ring
of $\Tw{M}/\ZZ_5$ is essentially the same as that of $\cal M$.  The
quotient of the quintic in $\IC^5$, $M/\ZZ_5$, which occurs in the
\LGO, is then seen as the orbifold limit of the blown-up quintic,
$\Tw{M/\ZZ_5}=\Tw{M}/\ZZ_5$, which is at the same time also a quintic
in ${\cal O}({-}5)$, which in turn contracts to $\cal M$ and so has
identical (co)homology and homotopy; see also Ref.~\rGRY. Since \CP4
itself is the exceptional divisor in ${\cal O}({-}5)$, its
intersection with the quintic (and a hyperplane) is the sought for
exceptional divisor corresponding to the ``K\"ahler twisted state''.

It is not hard to generalize this to hypersurfaces in weighted
\CP4's, but warped cases such as the one discussed earlier and
complete intersections in general pose a more difficult task.
Given a \CY\ complete intersection $\cal M$, we consider $\Tw{M}$, the
corresponding complete intersection in $\prod_r {\cal V}_r$, where
${\cal V}_r$ is the appropriate generalization of the total space of
${\cal O}({-}5)$ to the (weighted) factor \CP{n_r}. We next realize
that $\Tw{M}$ is a blow-up of $M$, the corresponding complete
intersection in $\IC^N$, where $N$ is the total dimension of the
product of the total spaces of ${\cal V}_r$. The (co)homology ring of
$\cal M$ is essentially the same as that of $\Tw{M}$ and the \LGO\
refers to $M$, the orbifold limit of $\Tw{M}$. Note also that $M$
looks like a (multi-)cone the (complex-projective) base of which is
$\cal M$ and that the action of $\Q$ is non-trivial only along the
generators. We then expect the twisted states to occur as the
intersection of the (multi-)cone and the exceptional divisor in
$\prod_r {\cal V}_r$.

\newsec{Final Remarks} 
In the foregoing analysis, we have been using (various) duality
relations quite often and rather freely; the careful reader might
have been worried whether this was always warranted and
well-defined. Let us here ward off such fears by noting that the
entire discussion is most appropriately set in the framework of the
Hilbert spaces, in which we focus on the subspace of massless
states, and which was implicitly the case throughout the article.
The scalar product there provides a perfectly canonical duality
relation allowing the freedom which we have utilized.

As we have mentioned, our analysis reveals a natural and manifest
isomorphism between the chiral and antichiral ring representatives
(both untwisted and twisted) as found via spectral flow arguments in
\LGO\ models~\rLGO\ and the tensor representatives found via Koszul
complex calculations.  An interesting question is to try to pursue
this correspondence to the level of ring structures. Our discussion
above makes it clear that there is a natural ring structure---a
Jacobian ring structure---which can isomorphically be imposed on the
mathematical cohomology and conformal field theoretic states.
Furthermore, Jacobian this ring structure exactly matches the one
supplied by the operator product expansion of conformal field theory
in the untwisted $(c,c)$ sector. It would be interesting to find some
direct method of calculating the conformal field theoretic ring
structure from the twisted chiral primary field representatives
studied here and in Refs.~\rLGO.

Given that elements of \hm{1}{{\rm End}\TM} can also be described by
the Koszul complex computation, our present analysis indicates that a
similar comparative derivation of these $E_6$ singlet fields is
possible. However, given the subtleties which we encountered in the
case of the simpler \hm{\star}{\TM} cohomology groups, we also expect
this comparison to be far from straightforward. These and other
related issues are being examined currently and we hope to report the
results in a subsequent article.

\vfill\bigskip
\noindent{\bf Acknowledgements.}\ \
We happily acknowledge the helpful discussions with
R.~Plesser, C.~Vafa and S.-T.~Yau.
P.B.\ has been supported by the Fulbright Program, the Robert
A.~Welch Foundation and the NSF Grant PHY9009850.
B.R.G.\ is supported in part by the National Science Foundation.
T.H.\ was supported by the DOE grant DE-FG02-88ER-25065.
\vfill
\eject

\listrefs

\bye